\documentclass[prl, reprint,  amsmath,amssymb, aps, longbibliography ]{revtex4-2}

\usepackage{graphicx}
\usepackage{dcolumn}
\usepackage{bm}
\usepackage[dvipsnames]{xcolor}
\usepackage{indentfirst}
\usepackage{hyperref}
\usepackage[mathlines]{lineno}
\usepackage[colorinlistoftodos]{todonotes}
\usepackage{physics}

\begin{document}

\title{Multi-loop and Multi-axis Atomtronic Sagnac Interferometry}

\author{Saurabh Pandey}
\altaffiliation[]{Present address: Sandia National Laboratories, 1515 Eubank Blvd SE, Albuquerque, NM 87123, USA}
\author{Ceren Uzun}
\author{Katarzyna A. Krzyzanowska}
\author{Malcolm G. Boshier}
\affiliation{Materials Physics and Applications Division, Los Alamos National Laboratory, Los Alamos, NM 87545, USA}

\date{\today}

\begin{abstract}
We report the experimental realization of a large-area and multi-axis atomtronic interferometer in an optical waveguide for rotation sensing. 
A large enclosed area is achieved through multi-loop operation in a guided atom interferometer using Bose-Einstein condensates. 
We demonstrate a three-loop interferometer with a total interrogation time of $ \sim$0.4\,s and an enclosed area of  8.7\,mm$^{2}$ —the largest reported in a fully guided or one-dimensional setup. 
High-contrast interference fringes are observed for up to five Sagnac orbits in a smaller loop-area configuration. 
Our approach enables interleaved rotation measurements about multiple arbitrary axes within the same experimental setup. 
We present results for area-enclosing interferometers in both horizontal and vertical planes, demonstrating that the interferometer contrast remains comparable across orthogonal orientations of the enclosed area vectors.
\end{abstract}

\maketitle

The interference of matter-waves is fundamental to quantum mechanics. 
The development of laser cooling, magnetic and optical trapping, and evaporation techniques has powered the field of atom interferometry to rapid growth.
Atom interferometry is typically done by coherently splitting a cloud of cold (ultracold) atoms into separate paths using light pulses, propagating them along different trajectories, and finally recombining them to generate interference fringes \cite{1966AltshulerPRL, 1986GouldPRL, 1988ClauserPBC, 1989BordePLA, MKasevich91PRL}. 
The phase of the fringe pattern is a direct measure of the phase difference accumulated by the wavepackets along their respective paths. 
The interferometer space-time geometry can be configured to perform precise measurements of inertial forces, fundamental constants, dark energy searches, gravitational wave detection, general relativity tests, and more \cite{2009CroninRMP, 2014BarrettCRP, 2018PezzeRMP, 2019BongsNRP, 2022NarducciAPX, Amico2022RMP, Amico2021AVS, Oh2024AIAA, Polo2024QST}.

Atom interferometry is implemented in two main settings: with freely falling cold and ultracold atoms \cite{MKasevich91PRL, Peters99Nature, Rosi14Nat, Parker18Sci, Stolzenberg25PRL, Halevy25arXiv}, or with ultracold atoms that are fully trapped  \cite{Hansel01PRA, 2024LeDesmaPRR, Xu19Sci, Duspayev21PRA, Navez16NJP, Dash24AVS, Prem2024PRA} or guided \cite{Wang2005PRL, Wu2007PRL, Henderson09NJP, 2020MoanPRL, McDonald2013PRA, kim2022Arxiv, Navez16NJP, Pandey2019Nat} in smooth optical and/or magnetic potentials. 
The latter, also known as the atomtronics approach, aims at developing tools to manipulate atomic matter-waves in controlled environments such as waveguides \cite{Amico2022RMP}. The advantage of this approach is that trapping the matter waves makes possible long interrogation times, and hence high sensitivity, in a physics package that can be much smaller than a system that must be large enough to accommodate atoms falling freely under gravity.
Our atomtronic Sagnac interferometer operates by splitting, reflecting, and recombining a Bose-Einstein condensate (BEC) using Bragg diffraction from a pulsed standing wave, with the BEC being guided within a linear optical waveguide throughout the process \cite{Krzy2023PRA}. 
To enclose an area, the guide is moved outward and back between the first and the last beam splitter pulse. 
A key distinction from other guided schemes is that we work with a delta-kick collimated \cite{ Ammann1997PRL, 2002ArnoldPRA, Pandey21PRL} large-size BEC containing only a few thousand atoms resulting in significantly decreased mean-field interactions. 
This enables us to push the interferometer interrogation time to hundreds of milliseconds while maintaining high fringe contrast. 
A practical approach to enclosing a larger geometrical area is working with multiple smaller loops. 
Area-enclosing multi- loop guided atom interferometers have been proposed \cite{Wu2007PRL, Moukouri2021arXiv} and experimentally studied with freely falling atoms \cite{Stockton2011PRL, Sidorenkov2020PRL}. 
With access to a longer interrogation time thanks to the delta-kick cooling, small BECs, and fringe phase correction using a classical accelerometer, we are able to increase the number of loops beyond one and extend the Sagnac area by more than one order compared to our previous generation interferometer \cite{Krzy2023PRA}.
In a two-dimensional magnetic trap (time-orbiting potential trap), Beydler \textit{et al.} have recently achieved a two orbit BEC interferometer enclosing 8\,mm$^{2}$ Sagnac area \cite{Beydler2024AQS}. 
A unique and powerful feature of our scheme is that by moving the guide in different planes (keeping the waveguide’s propagation direction normal to gravity), it is easily possible to sense rotation rates about multiple axes without the issue of wavepacket trajectories being impacted by the gravity. 
This is not straightforward to implement in other guided schemes, where gravity compensation would be needed when the loop plane is tilted away from horizontal \cite{Beydler2024AQS}.

In this paper, we present a large-area, multi-loop waveguide atom interferometer designed for rotation sensing about multiple axes. 
The largest enclosed area achieved in our experiments is 8.7\,$\text{mm}^2$ with three Sagnac orbits with an orbit time of 124\,ms per loop. Additionally, we demonstrate a five-loop interferometer enclosing a smaller area of 0.13\,$\text{mm}^2$ with a shorter orbit time of 22\,ms per loop. 
To demonstrate multi-axis rotating sensing, we explore two configurations: one with the waveguide oriented in the vertical plane (large-area, multi-loop results), and another in the horizontal plane. Both configurations produced comparable fringe contrasts for similar enclosed areas, demonstrating the capability to sense in orthogonal orientations.

\begin{figure}[h]
\includegraphics[width=0.45\textwidth]{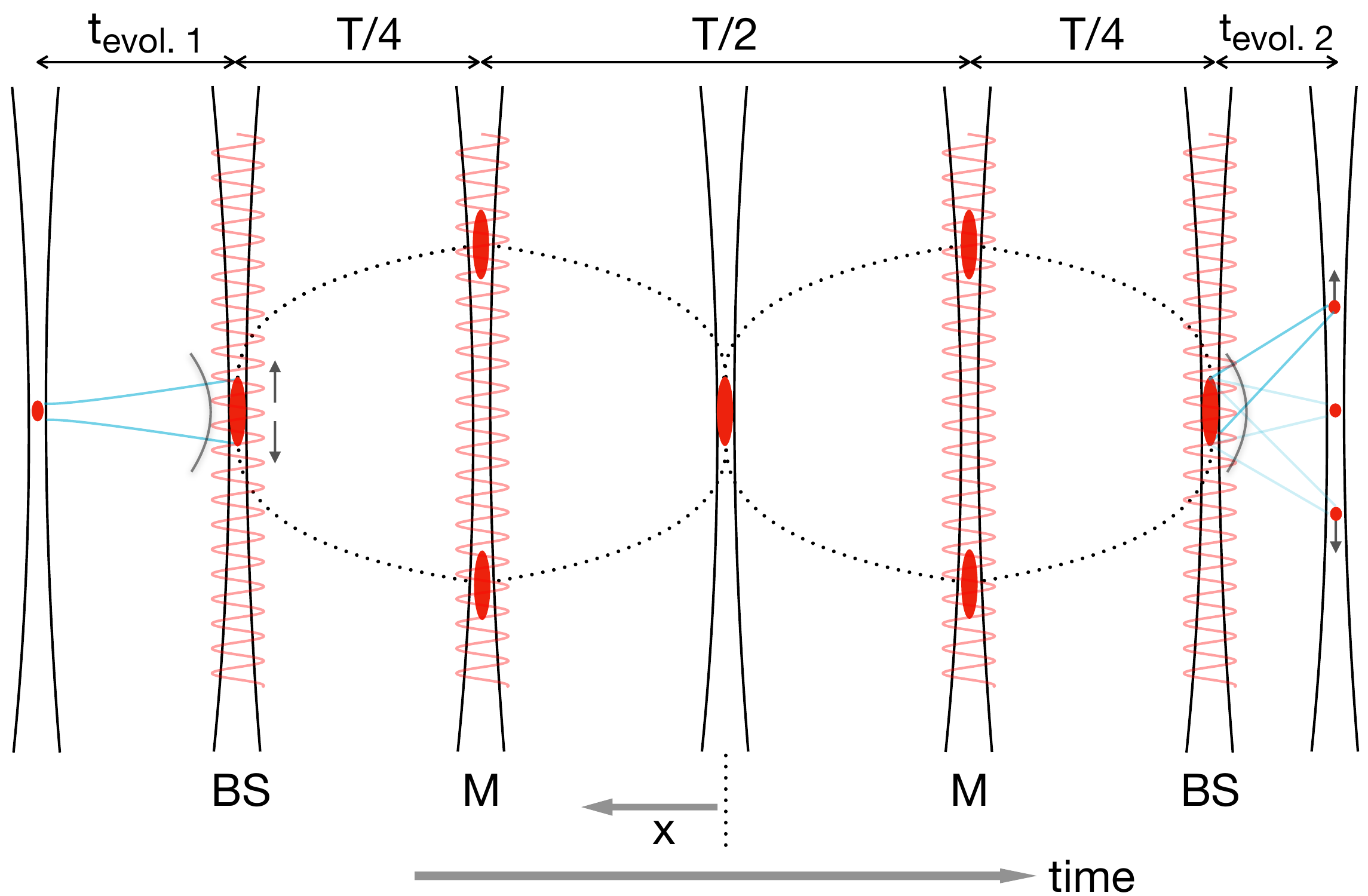}
\caption{One-loop moving waveguide unfolded interferometer scheme. 
The sequence starts by releasing a BEC from a crossed dipole trap into the guide. 
After an expansion duration of $t_\text{{evol.1}}$, a delta-kick collimation lens is applied, and right after that a beam splitter (BS) pulse is shone on a collimated wave-packet. 
After a guide moving time of $T/4$, a mirror (M) pulse is shone that reverses the sign of the atomic momentum. 
At the mid point of the interferometer at $T/2$, the guide motion is reversed. 
A second M pulse is applied at $3T/4$. 
The wave-packets are recombined at $T$ with a second BS at their starting position. 
Right after the final BS, a delta-kick focusing lens is applied that tightly focuses all the three momentum states in $t_{\text{evol.2}}$, and the three states separate spatially. 
The dotted curves show the trajectory of the atoms (see main text).
}
\label{fig: schemeInterferometer}
\end{figure}

The experimental setup is a more recent version of the one described in reference \cite{Krzy2023PRA}. 
We generate a BEC of approximately 1000  $^{87}$Rb  atoms in the $|1,\,-1\!>$ state using a crossed optical dipole trap with 1064\,nm beams. 
The horizontal and vertical beams have waists of $17\mu$m and $88\mu$m, respectively. 
The atoms are then transferred to the first-order magnetically insensitive  $|1,\,0\!>$ state via a Landau-Zener RF sweep followed by a magnetic gradient kick to remove unwanted $F=1$ states (see SM1). 
The BEC is adiabatically loaded into another crossed dipole trap formed by a horizontally oriented guide beam (waist $67\mu$m) and the same vertical beam. 
The guide trap frequencies are measured as $(\omega_\text{x}, \omega_\text{y}, \omega_\text{z}) \approx 2\pi \times (140, 140, 0.5)$\,Hz, while the vertical beam has a transverse trap frequency of 20.3\,Hz. 
The BEC is loaded into the guide by ramping down the vertical beam power and expanded along the guide at 1.54 mm/s. 
After a 60\,ms expansion, we apply a delta-kick "lens" to collimate the wavepacket, achieving a final effective temperature of 1.1\,nK and an axial size of $120\mu$m (Thomas-Fermi FWHM) in the guide. 
To enable this collimation, the vertical beam is time-averaged (painted) at 10\,kHz to create a large harmonic lens of FWHM size $180\mu$m. 
This painted lens is also utilized to focus the wavepackets at the interferometer readout (see Fig. 1). 
The measured trap frequency of the painted lens is 19.7\,Hz. 
The delta-kick collimation and focusing lens durations are 0.9\,ms and 2\,ms, respectively.

The interferometer sequence begins right after the delta-kick pulse prepares a collimated wavepacket. 
Fig. \ref{fig: schemeInterferometer} explains the sequence of our moving guide interferometer \cite{Krzy2023PRA}. 
The guide translation is performed by a one-axis acousto-optic deflector (AOD, IntraAction ATD-274HD6), placed one focal length ($f$) behind a $f=20$\,cm plano-convex lens. 
The focus of this lens is aligned to the trap inside the science cell. 
By changing the RF frequency to the AOD, the beam after the lens is translated by $338\mu$m for 1\,MHz change in the RF frequency. 
The beamsplitters and mirrors of the atom interferometer are implemented through Bragg scattering off a near resonant standing wave formed by retroreflection from a dichroic mirror placed in the guide beam  \cite{Krzy2023PRA}.   
A double-square intensity pulse is used as a symmetric beam splitter \cite{Wu2005PRA} and a Gaussian pulse for the mirror operation on BECs \cite{Mueller2008PRA}. We measure nearly 100\% splitting and reflection pulse efficiency.
We employ  $\pm 2\hbar k$ (or 1.2\,cm/s for $^{87}$Rb with D2 line) momentum transfer, where $k=2\pi/780$\,nm. 
The beamsplitter pulse is composed of two square pulses of  22.3\,$\mu$s and 21.2\,$\mu$s with a no-light gap of 37.9\,$\mu$s in between. 
The Gaussian reflection pulse has a FWHM duration of 44.3\,$\mu$s. 
The Bragg laser is $\approx$ 12.6\,GHz blue-detuned from the  $5S_{1/2},\,F_\text{g} = 1 \xrightarrow{} 5P_{3/2},\,F'$ transition (see SM2 on Bragg setup). 
The one-loop interferometer uses a Bragg pulse sequence of beamsplitter-mirror-mirror-beamsplitter. 
To scan the interferometer phase, the Bragg laser frequency is varied by applying a voltage to the piezo of the laser at the recombination stage (more details in SM3). 
Right after the recombination pulse, a strong delta-kick pulse is applied that focuses all the three momentum states to a tight spot in 20\,ms, at which point the states are separated by  260\,$\mu$m. 
This final lens drastically improves the signal to noise ratio, which is particularly important given the low atom number and large size of the BECs. 
We use absorption imaging to count the number of atoms in the three peaks with no time of flight. 
Final analysis is performed on the absorption images after applying a fringe-removal algorithm \cite{Li2007COL, Ockeloen2010PRA}.

We can easily switch between static and moving guide interferometers in the experiment by either keeping the RF frequency to the AOD fixed (static beam) or ramped (moving case). 
A static guide interferometer does not enclose any area, but its performance gives us a reference point for the moving guide case. 
The guide is moved using the shortcuts to adiabaticity (STA) protocol \cite{Torrontegui2011PRA}. 
The enclosed area is approximately 25\,$\%$ larger with the STA trajectory compared to moving the guide at a constant velocity (see SM4). 
This is because the guide, starting from rest, moves more slowly at the beginning, turning point(s), and end of the loop. 
The parameters that define the RF frequency ramp to move the guide include the transverse trap frequency, the desired guide translation, and the time required to traverse the interferometer. 
To enable fast transport, the transverse confinement is kept strong, typically operate above 100\,Hz. 
Meanwhile, the static guide can be operated at much lower power providing just enough confinement to prevent atoms from free fall.

A challenge associated with interferometers that use a mirror to retro-reflect a laser beam to create the Bragg pulses for the splitting and mirror operation on atoms is the vibration of the mirror itself. 
This makes the phase of the standing wave light pulses different than programmed and injects noise into the interferometer. 
We use a classical accelerometer to measure the mirror acceleration during the interferometer sequence and post-correct each run’s phase to remove this effect \cite{Leduc2003JOB, Gouet2008APB, Barrett2014PISP}. 
The change in the interferometer phase due to the mirror motion is calculated by multiplying the measured mirror acceleration $a(t)$ by the double triangle like response function $f(t)$ and integrating the product over the interferometer cycle, $\Delta \Phi = 4 n k \int_{}^{} f(t) a(t) \text{d}t$, where integer $n$ is the diffraction order ($n=1$ for $2\hbar k$).

\begin{figure}[t]
\includegraphics[width=0.48\textwidth]{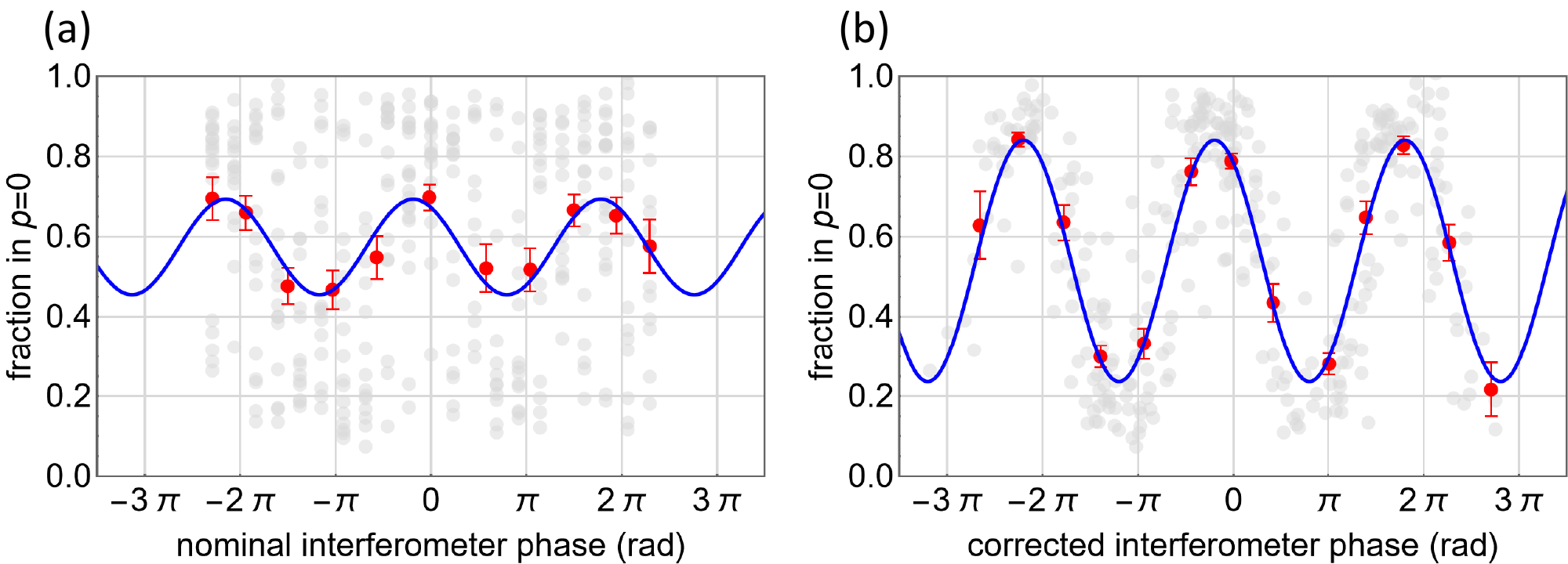}
\caption{Accelerometer phase correction. 
Improvement in the average fringe contrast after applying the phase correction to the individual runs of a 121\,ms one-loop static guide interferometer.
 a) Without phase correction and b) Phase corrected fringe. 
The gray dots are individual experimental data points, the red points are binned averages of those measurements, and the blue curves are least-squares fit of a cosine function to the average points.
}
\label{fig: accelFringeCorr}
\end{figure}

We have installed two navigation grade accelerometers (InnaLabs AI-Q-2110) in the experiment to compute the phase correction. 
The first one is attached to the retro-reflection mirror mount and the second to the experiment breadboard to which all the optic posts are mounted. 
The phase correction given by the two accelerometers agree well. 
There is a factor of roughly two times average contrast improvement after applying the phase correction. 
This method of accelerometer correction works equally well for a static or moving guide interferometer and depends only on $T$. 
Fig\,\ref{fig: accelFringeCorr} shows $T = 121$\,ms one-loop static guide fringes, without (a) and with (b) the accelerometer phase correction. 
The peak-to-peak fitted fringe contrast increased from 0.23 to 0.6 by applying the correction.
Here, the guide axial confinement is weaker $(\omega_\text{z}/2\pi \approx 0.4\,\text{Hz})$ than the moving guide cases in the results below.

We quantify the fringes by least-squares fitting the $\textit{p}=0$ fraction data, $N_{0\hbar k}/(N_{-2\hbar k}+N_{0\hbar k}+N_{+2\hbar k})$, to the function $P(\phi)= \textit{O} + \textit{C}/2\,\text{cos}(\phi + \phi_{0})$, where \textit{O} is the fringe offset, \textit{C} is the fringe contrast, $\phi$ and $\phi_{0}$ are the interferometer and reference phase, respectively. We obtain an interference fringe by scanning $\phi_{0}$ (more details in SM3).

In our configuration, each wave-packet completes an integer number of full orbits around the Saganc loop, with the final beamsplitter readout occurring at the same point as the initial splitting. As a result, the interferometer is sensitive to rotation and but not to constant acceleration.
The interferometer phase shift from the rotation of the apparatus (or Earth) is $\Phi_\text{S}=4 \pi \boldsymbol{\Omega \cdot \textbf{A}}/(h/m)$, where \textbf{A} is the loop area, $\boldsymbol{{\Omega}}$ is the rotation angular velocity, and $m$ is the atomic mass. 
It is of course critical to identify and minimize any systematic phase errors.

A typical calibration of the interferometer sequence involves a scan of the mirror to mirror pulse time to maximize the fringe contrast by maximizing the overlap of the wavepackets when the recombination pulse is applied. 
The width of the overlap time window depends on the size of the wave-packets and the potential curvature. We typically measure a FWHM of about 0.3\,ms. 
It is worth noting that a higher axial guide curvature or interferometer $T$ both shorten the time wave-packets take to overlap again at the end \cite{Burke2008PRA, Stickney2007PRA}.
The deviation in $T/2$ ("mirror to mirror time") in a harmonic potential from being in a flat potential for the overlap condition  is $\,\omega^{2}\,T^{3}/32$, where $\omega$ is the axial trapping frequency of the guide. 
The experiment follows this dependence very well (see SM5).
 
\begin{figure*}[bt]
\includegraphics[width=1\textwidth]{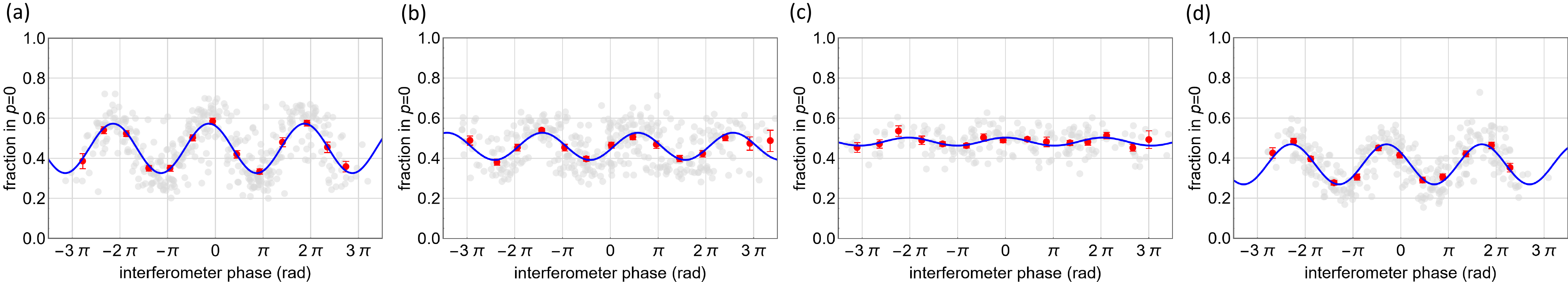}
\caption{Multi-loop moving guide interferometry.
 a) One-loop fringe for $T$ = 124\,ms and 2.9\,mm$^2$ enclosed area.
b)  Two-loop fringe for a total $T$ = 246\,ms and 5.8\,mm$^2$ enclosed area.
c) 371\,ms total $T$ three-loop fringe with of 8.7\,mm$^2$ enclosed area.
d) A five-loop fringe for a total $T$ of 108\,ms and 0.13\,mm$^2$ enclosed area.
The guide was moved vertically in all these interferometers. 
The red (gray) points show the bin-average (raw) fringe data. The blue curve is the cosine fit.
}
\label{fig: multiLoop}
\end{figure*}

To increase the number of loops, we turn off the last beam splitter pulse, allow the wavepackets to go through each other, and repeat the guide motion and mirror pulse sequence.
The second beam splitter pulse is applied when the atoms come back to the starting position after an additional even number of mirror pulses.
Due to a restriction of the optical system on how far the guide beam can be moved, we have slightly modified the loop geometry (single and multiple) to further increase the enclosed area. This is done by keeping the guide static for a short time in the beginning, the guide turning points, and end. In these regions, the guide is at rest or moves slowly. This allows the wave-packets to move farther along the guide, enlarging the loop area.
This $t_\text{hold}$ is placed symmetrically, $t_\text{hold}/4$ at the beginning, $t_\text{hold}/2$ in the middle, and $t_\text{hold}/4$ at the end of the interferometer. 
This short hold in the static guide increases the Sagnac area for any number of loops, for e.g., from 2.7 to 2.9\,mm$^2$ for a one-loop interferometer shown in Fig.\,\ref{fig: multiLoop}a.

Fig.\,\ref{fig: multiLoop} shows our multi-loop and large-area interferometer experiment results. 
In all the fringes discussed below, the radial trapping of the guide was 140\,Hz.
In the first three fringes of Fig.\,\ref{fig: multiLoop}, a to c, the guide was moved by 2.7\,mm in 50\,ms vertically downward and then brought back to the starting position in 50\,ms. 
The inclusion of hold intervals where the guide position was held fixed increased the single-loop time to the values given below. 
All the fringes presented below are phase corrected from the accelerometer. 
Fig.\,\ref{fig: multiLoop}a shows a contrast $C=0.26$ fringe for one-loop moving guide interferometer with $T$ = 124\,ms and enclosed area of 2.9\,mm$^2$. The guide was static for 24\,ms and moved for 100\,ms in total.
For the two and three-loop fringes, the guide was static for 22\,ms and moved for 100\,ms in each loop.
Fig.\,\ref{fig: multiLoop}b shows a two-loop  contrast $C=0.13$ fringe with $T$ = 246\,ms and 5.8\,mm$^2$ enclosed area. 
The three-loop interferometer experiment was a direct extension of the two-loop one, see Fig.\,\ref{fig: multiLoop}c. We set $T = 371$\,ms and the computed enclosed area is 8.7\,mm$^2$. The fitted fringe contrast is  $C=0.05$ 
Fig.\,\ref{fig: multiLoop}d is a five-loop fringe (contrast $C=0.2$) of total interrogation time of 108\,ms. The $t_\text{hold}$ was set to zero for this five-loop configuration. The guide translation here was smaller, 0.17\,mm.

\begin{figure}[h]
\includegraphics[width=0.48\textwidth]{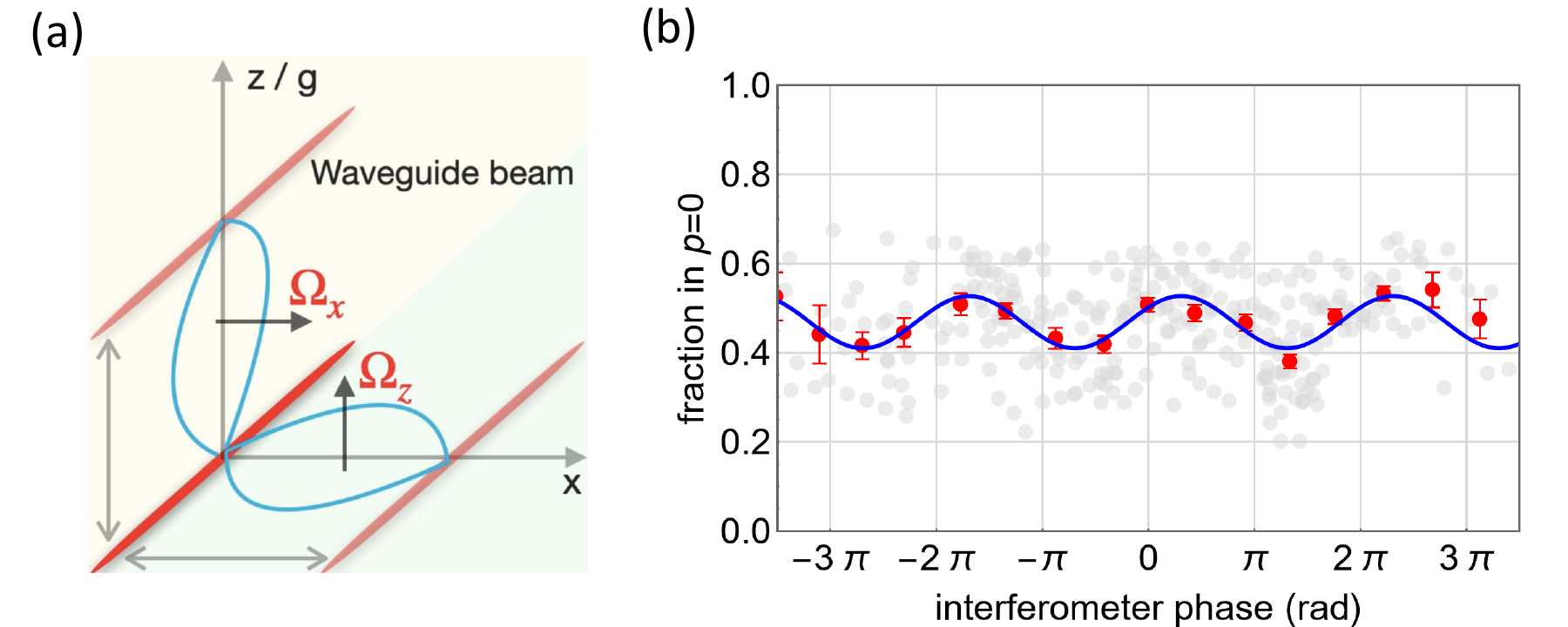}
\caption{Multi-axis interferometry. 
a) Waveguide moving in the horizontal and vertical plane forming two Sagnac loops with orthogonal area vectors for sensing $\Omega_\text{x}$ and $\Omega_\text{z}$. 
b) One-loop interferometer fringe for the guide moving in the horizontal plane. 
The interrogation time and enclosed area is 160\,ms and 3.2\,mm$^2$, respectively. 
}
\label{fig: horMoving}
\end{figure}

Our interferometer scheme has the potential to sense rotation rates about multiple arbitrary axes without any hardware changes in the experiment. 
We currently use a one-axis AOD that deflects the guide only in the vertical plane.
All the fringes shown in Fig.\,\ref{fig: multiLoop} are with the vertically moving guide.
By rotating the AOD by 90$^{\circ}$, we performed interferometry where the guide carrying atoms moves in the horizontal plane. 
Fig.\,\ref{fig: horMoving} shows a fringe in this configuration for an interrogation time of 160\,ms. 
The total hold in the static guide was 80\,ms, moved for 80\,ms, and the guide translation was 2\,mm. The wavepackets enclose an area of 3.2\,mm$^2$ and  the fitted fringe contrast is $C=0.12$ 
We measure similar fringe contrast for the vertically and horizontally moving guide interferometers when the guide translation and interrogation time are kept the same. 
A comparison can be drawn between the vertical loop interferometer with area 2.9\,mm$^2$ (Fig.\,\ref{fig: multiLoop}a) and the horizontal loop enclosing 3.2\,mm$^2$ area (Fig.\,\ref{fig: horMoving}b).

We now discuss the factors that can limit the performance of our interferometer.
The fringe underlying contrast is affected by factors such as overlap of the wave-packets at the end, non-optimal Bragg pulse intensity, imperfect overlap of the Bragg and guide beams, experiment table floating condition, the Zeeman state of the atoms, the axial curvature of the guide, and the starting position of the wave-packets with respect to the waist of the guide beam. 
We do measure a significantly lower underlying fringe contrast when the interferometer is operated with the initial location of the wavepacket located away from the guide waist and/or with high axial curvature \cite{Burke2008PRA, Horikoshi2006PRA}. 
The axial trap frequency sets the size of the collimated BEC (to be the ground state size) that should ideally be used in the interferometer. Otherwise, the long-axis size after the collimation lens oscillates and degrades the interferometer contrast.

For long interrogation times, a small amplitude and slow oscillation of the table can be picked up by the atoms as an initial velocity and can result in population imbalance between the two interferometer arms and improper overlap at the end. We noticed this for $T$=120\,ms and longer.
We are currently investigating the initial velocity fluctuation of the condensate, which should ideally be much smaller compared to $\pm 2\hbar k$ or 11.8\,mm/s. 
Overall, we believe that an improved condensate collimation and initial velocity stability should result in a higher underlying and averaged fringe contrast, respectively.

We noticed that the waveguide waist moved for a duration of a few seconds due to the thermal lensing caused by the AOD. Since the total duration of the longest interferometer we ran is 400ms, it is a factor that can affect the wave-packets dynamics. A better solution will be to use a piezo-based fast-tip/tilt mirror. A piezo mirror also offers a large deflection range and will not cause any disturbance to the waveguide profile.

Lastly, the phase diffusion time due to the atomic interactions is estimated to be about 1\,s for our system (for the phase noise to be $\pi$) \cite{Fallon2015JPB}. It is not a leading cause of the reduced average contrast here.  

Overcoming the loss of fringe contrast at longer interrogation times and using a larger deflection two-axis AOD or piezo based tip/tilt mirror would allow us to study bigger Sagnac loops. 
Routing the vertically deflected guide beam back to the trap location would be a straightforward way to realize an interleaved three orthogonal Sagnac atom interferometers.
Another promising direction to increase the loop area is working with higher momentum wave-packets \cite{Giltner1995PRL, McGuirk2000PRL, Mueller2008PRL, Hughes2007PRA}. 
We are currently investigating higher momentum guided-atom interferometers using novel pulse shapes to realize high fidelity and intensity-robust beam splitters \cite{Clare2021JAP}.

Angle random walk (ARW, in deg/$\sqrt{\text{hr}}$) and bias stability (in deg/hr) are the two most important parameters of a gyroscope. ARW depends on the noise in the fringe phase (represents sensitivity), and bias stability is related to drift in the averaged phase (any offset from the true rotation).
With the largest Sagnac area reported above, 8.7\,$\text{mm}^2$, our interferometer has a scale factor ($\Phi_\text{S}/\Omega=2mA/\hbar$)  of 2.4 x $10^4$ rad/(rad/s). With the fit phase uncertainty of 0.24\,rad, we estimate a rotation error (phase uncertainty/scale factor) of 20\,$\mu$rad/s. With a total measurement time of 1.35 hours, we estimate an ARW of 2.4 deg/$\sqrt{\text{hr}}$, comparable to a recent demonstration in a two-dimensional waveguide atom interferometer \cite{Beydler2024AQS}. We would like to emphasize that currently, with an increase in the number of loops or Sagnac area, contrast goes down due to technical (not fundamental) issues, and ARW remains of that order as mentioned above. 

We have observed high underlying but poor average contrast in a one-loop static guide interferometer for $T$ up to 400\,ms. 
For 1\,cm guide translation and $6 \hbar k$ splitting, a 400\,ms one-loop interferometer will give an area of 1\,cm$^2$ translating to an order higher scale factor of $\sim$ 3 x $10^5$\,rad/(rad/s). A sub-Hz BEC producing machine would be desirable to further improve the performance \cite{Vendeiro22PRR}.
We measured the Allan Deviation of the interferometer phase, and it follows the $1/\sqrt{\tau}$ scaling up to $10^{4}$ s \cite{Krzy2023PRA}. 
Since gyro's bias stability is one of the most important parameters together with the sensitivity, we believe our approach has potential to meet the navigation level performance.  

In summary, we have experimentally realized a multi-loop and multi-axis guided atom interferometer. 
The Sagnac area reported here is the largest realized in a waveguide so far.
To further validate the promise of a multi-loop approach, we presented a high-contrast five-loop moving waveguide interferometer for a smaller loop area. 
In addition, we offer a solution to the challenging and important problem of multi-axis guided atom interferometry. 
We presented similar performance interferometers with Sagnac loops oriented in the horizontal and vertical plane. 
Our results on guided atom interferometry are an important step towards realizing a compact atomic gyroscope for applications such as inertial navigation.

The data that support the findings of this study are available from the corresponding author upon reasonable request.

The authors declare no competing interests.

This work was funded by DARPA under the A-PhI program. The initial development of the experimental apparatus was supported by the Office of Naval Research and by the Laboratory Directed Research and Development program of Los Alamos National Laboratory under project number 20180045DR.

\clearpage

\section*{Supplemental Material} \label{SM}
\subsection*{SM1. BEC production, loading into the waveguide, and guide alignment}

The experiment begins by loading $^{87}$Rb atoms into a three-dimensional magneto-optical trap (3DMOT) from another 3DMOT via push beam. 
MOT loading is typically done in 5\,s. 
The atom cloud is then compressed and cooled during the compressed MOT (CMOT) and molasses stage, respectively. 
At the end of molasses, we switch off the repumper beam transferring the atoms from $F = 2$ to $F = 1$ hyperfine state in $5S_{1/2}$. 
Next, we capture $\sim$ $5\times10^{6}$ atoms in the $|1,\,-1\!>$ state at a temperature of $4\,\mu$K into the magnetic quadrupole trap from molasses. 
The cloud is further compressed by ramping up the magnetic gradient in the presence of crossed dipole trap beams. 
In our setup, we have three optical dipole beams at 1064\,nm wavelength, two co-propagating in the horizontal plane and one in the vertical direction (gravity axis). 
The two horizontal co-propagating beams have waists of 17\,$\mu$m ($b_\text{x1}$) and 67\,$\mu$m ($b_\text{x2}$). 
The waist of the vertical beam ($b_\text{z}$) is 88\,$\mu$m. 
We load about $7\times10^4$ atoms at $0.5\,\mu$K in a crossed dipole trap by ramping down the magnetic gradient to zero in two sequential exponential ramps. 
We also ramp down the dipole beam power during the loading stage. 
BECs are produced by evaporating the atom cloud in a crossed dipole trap of $b_\text{x1}$ and $b_\text{z}$. 
We routinely get pure BECs containing up to $6\times10^{3}$ atoms in $|1,\,-1\!>$ state.

We now describe the steps to prepare the atoms in a magnetically insensitive $|1,\,0\!>$ state and then loading them into the guide.
With the BEC held in a crossed dipole trap of $b_\text{x1}$ and $b_\text{z}$, we shine an RF field with its frequency sweeping from 950\,kHz $\xrightarrow{}$ 450\,kHz in 10\,ms, in the presence of a 1\,G homogeneous magnetic field. 
This transfers the atoms from $|1,\,-1\!>$ state to other two Zeeman sub-states of $F = 1$ $(m_\text{F}=0\,\&\,m_\text{F}=1)$. 
The magnetically sensitive states are filtered from the guide by applying a magnetic quadrupole gradient kick for 10\,ms.
Finally, we linearly ramp up the guide beam ($b_\text{x2}$) power to 420\,mW, in 200\,ms and adiabatically transfer up to $\sim$ 3000 Bose condensed atoms by ramping down ($b_\text{x1}$) and ($b_\text{z}$) beam powers to zero in 200\,ms.
One experiment cycle takes around 12\,s. 
The guide beam power is kept constant during the interferometer sequence. 
All three optical dipole beams are intensity stabilized using photodiodes and closed-loop feedback technique. 

We characterize the guide alignment with BECs. 
After loading the BEC in a crossed dipole trap of guide and the vertical beam, and then turning off the vertical beam fast we let the BEC go through shape and center of mass oscillation in the guide.
We use the long-axis center of mass oscillation of the BEC to precisely tune the guide levelling with respect to the gravity. 
This method enables us to align the position of vertical dipole beam to the waist of the guide and thus load the atoms at the bottom of the potential. 
The procedure is iterated between high and low guide axial curvature to decouple guide levelling from the misaligned guide waist. 
A long probe time of up to 10\,s allows us to get rid of any center of mass motion in a sub-hertz confinement trap. 

\subsection*{SM2. Bragg beam setup}
\begin{figure}[h]
\includegraphics[width=0.47\textwidth]{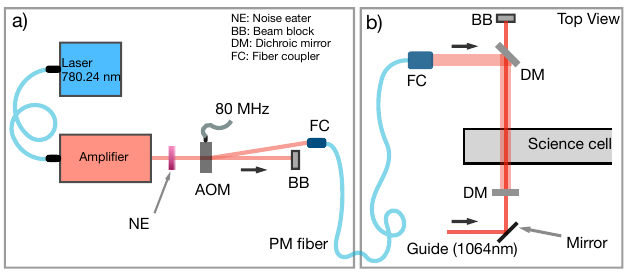}
\caption{Bragg beam setup. a) Bragg laser and AOM control. b) Relevant components of the BEC experiment setup. 
}
\label{fig: braggSetup}
\end{figure}

Our Bragg beam setup scheme is shown in Fig.\,\ref{fig: braggSetup}. 
It mainly consists of a laser (New Focus Vortex Plus TLB-6813-P) centered at 780.24\,nm wavelength, an amplifier (New Focus TA-7613), an acousto-optic modulator (Isomet 1206C-1), a noise eater (Thorlabs NEL03), and a few optical elements to focus and guide the laser beam. 
The laser is $\approx$ 12.6\,GHz blue-detuned with respect to the $^{87}$Rb $5S_{1/2},\,F_\text{g} = 1 \xrightarrow{} 5P_{3/2},\,F'$ transition. The laser is free-running with a center wavelength at 780.3 nm.
The output of the laser is fiber-coupled to the amplifier. 
The amplifier output goes through a noise eater for power stabilization and allows us to change the power level. 
Noise eater output is directed to an AOM operating at 80\,MHz and its first order diffraction is coupled into the fiber.
At the fiber exit, the beam is collimated to a $1/e^{2}$ waist diameter of 5.42\,mm using a Thorlabs TC25APC-780 triplet collimator. 
The beam is then reflected with a dichroic mirror towards the science cell. 
At the other end of the cell, the beam is incident normally on a second dichroic mirror, is reflected, and thus creates the standing-wave for the BEC splitting and reflection operation. 
The 1064\,nm guide beam passes through the dichroic mirrors and is overlapped with the 780\,nm standing-wave beam.

\subsection*{SM3. Interferometer phase $\&$ Bragg laser frequency stability}
The phase of our atom interferometer is, $\Phi = 2 n (\phi_1 - 2 \phi_2 + 2 \phi_3 - \phi_4)$, where $\phi_i$ is the phase of $i^\text{th}$ Bragg pulse with momentum $p= \pm 2 n \hbar k$ ($n=1$ in this work).
We scan the interferometer phase by changing the Bragg laser frequency at the recombination stage (last pulse). 
This is done by changing the piezo voltage in the Bragg laser. 
Changing the laser frequency by $\Delta f$ at the last pulse changes the phase of the pulse by $\Delta \phi_{4} = (4 \pi L/c) \Delta f$, where $L$ is the distance between the BEC and the retroreflecting mirror. 
Typically, we get a fringe by scanning the piezo voltage from -2 to 2\,V (detuning scanned from 11.8 to 13.4\,GHz, $5S_{1/2},\,F_\text{g} = 1$ to  $5P_{3/2},\,F'$). 
In each experimental run, the piezo voltage is ramped to a desired value between the second mirror pulse and the recombination pulse.
To avoid hysteresis issues, we then ramp the piezo back to -2\,V right after the interferometer cycle is over and it stays there. 
To scan the interferometer fringe over $2 \pi$, the laser frequency is changed by $\approx$ 0.64\,GHz.
This calculation is consistent with the separation between the retro-mirror and trap location of about 12\,cm.  

We did some tests to characterize the long-term (hours timescale) Bragg laser frequency stability and reproducibility when the piezo voltage is scanned. 
First, by just measuring the Bragg laser frequency with a wavelength meter. 
Second, we measured the BEC splitting fidelity at different piezo voltages between -2 and 2V. 
Lastly, we measured the transmitted intensity of the Bragg beam passing through a Rubidium vapor cell, on a photodiode. 
By changing the piezo voltage, we can tune the absorption (laser frequency tuned around the D2 line) and thus measure the frequency stability and reproducibility. 
All of those tests confirmed that the laser frequency (detuning) is stable at about 100 MHz level and reproducible for our current interferometer setting.

\subsection*{SM4. Guide \& wave-packets trajectory in a moving guide interferometer}
\begin{figure}[h]
\includegraphics[width=0.38\textwidth]{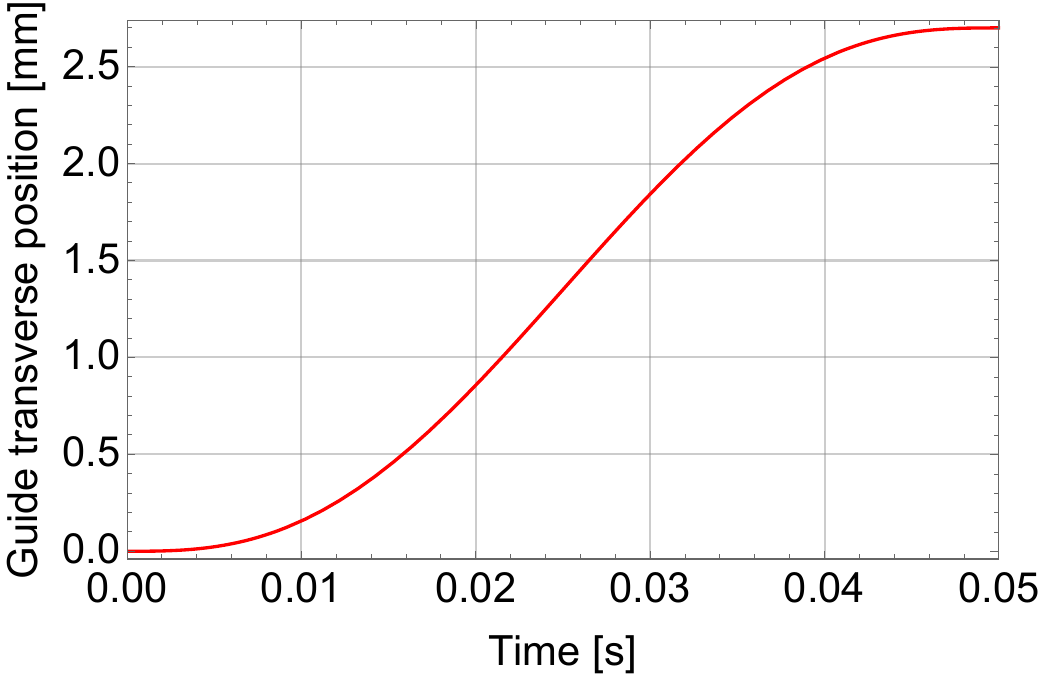}
\caption{Guide transverse position as a function of time. 
}
\label{fig: sta}
\end{figure}

We follow the shortcut to adiabaticity (STA) protocol to move the guide during the interferometer \cite{Torrontegui2011PRA}.
The guide's transverse trajectory depends on the translation distance, duration, and its transverse trapping frequency. 
The radial trapping frequency was fixed to 140\,Hz in all the configurations presented in this work.
Fig.\,\ref{fig: sta}) shows the trajectory of the guide for 2.7\,mm translation in 50\,ms. 
This configuration was used in the single and multi-loop interferometers of Fig.\,\ref{fig: multiLoop} in the main text. 
The plot only shows the guide motion in the first half of the interferometer cycle.
 In the second half, the guide is moved back to the original position following the same trajectory.

\begin{figure}[h]
\includegraphics[width=0.48\textwidth]{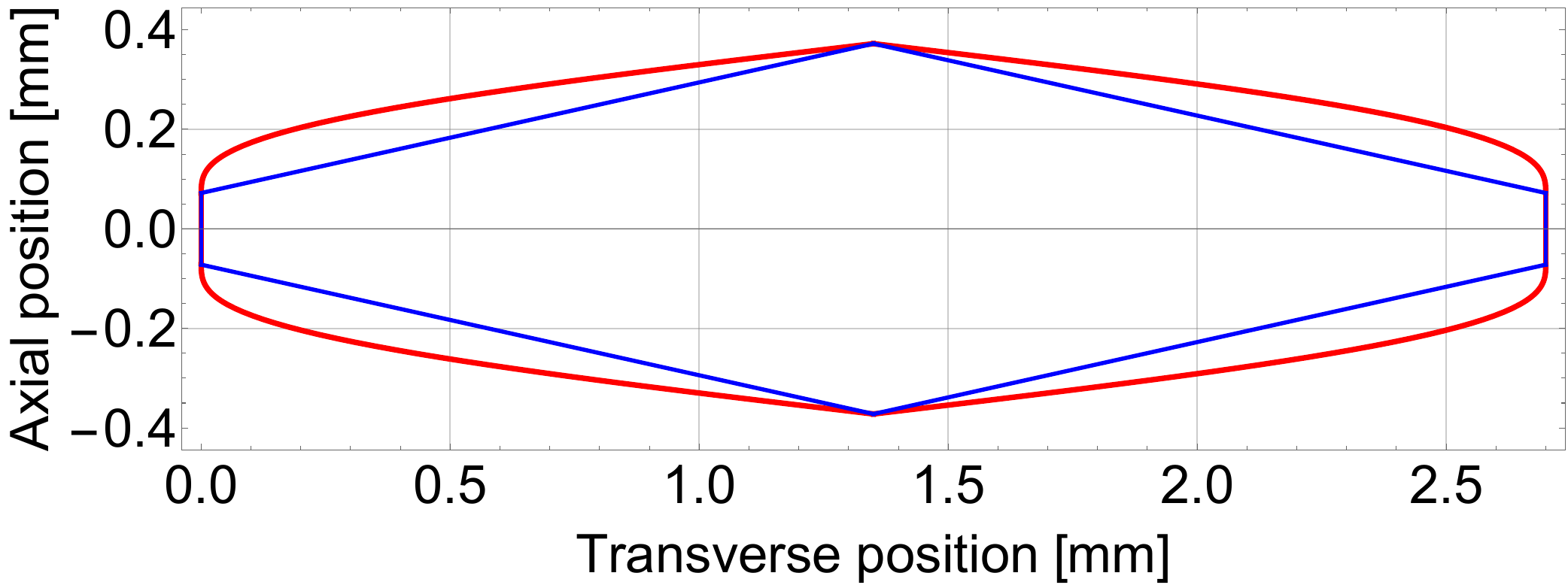}
\caption{One-loop moving waveguide interferometer geometry. 
This is for an interrogation time of 124\,ms. 
The guide is moved by 2.7\,mm in 50\,ms. 
There is a 6\,ms hold in the guide in the beginning and at the turning point on the right hand side. 
The red curve is with STA and the blue curve is if the guide was moved with a constant velocity.
}
\label{fig: loopGeometry}
\end{figure}

Fig.\,\ref{fig: loopGeometry} shows the trajectory of the two wave-packets (+$2 \hbar k$ and -$2 \hbar k$) during the interferometer. 
The red curve is with STA, and the blue curve shows if the guide was moved with a constant velocity. 
It is clear that with STA the wave-packets enclose more area. 
This is due to the fact that the guide moves slowly in the beginning and at the end, and thus the wave-packets move farther along the guide in those regions. 
This parametric plot is with a short hold time in the beginning and at the end of the guide motion. 
Fig.\,\ref{fig: loopGeometry} shows the loop that was implemented in the one-loop interferometer experiment presented in Fig.\,\ref{fig: multiLoop}a ($T$ = 124\,ms and 2.9\,mm$^2$ enclosed area).  

\subsection*{SM5. Wavepacket overlap time versus guide axial curvature \& $T$}
\begin{figure}[h]
\includegraphics[width=0.4\textwidth]{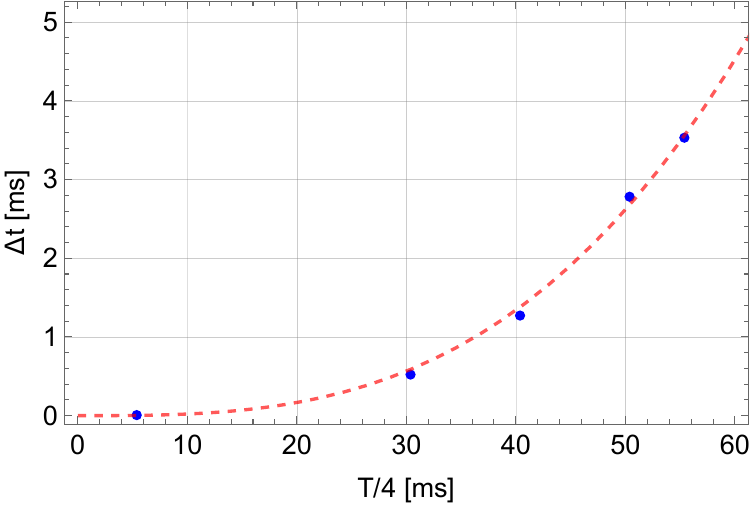}
\caption{The difference in the experimentally measured mirror to mirror time from the exact $T/2$ values versus one-fourth of the interferometer interrogation times. Blue dots are the data points, and the red dashed curve is a fit to them.}
\label{fig: overlapTiming}
\end{figure}

We scan the duration between the first and the second reflection pulse in the interferometer to find the maximum overlap of the wave-packets at the end and thus maximize the fringe contrast.
In this way, we keep the interferometer symmetric.
The correction to the return time of the wave-packet at the recombination pulse is $\omega^2 t^3$ \cite{Burke2008PRA}, where $\omega$ is the guide axial trapping frequency and $t$ is the $T/4$ time (splitting to reflection). 
The required change in the mirror to mirror time interval is twice that amount, or 2$\omega^2 t^3$.
Fig.\,\ref{fig: overlapTiming} shows the measured deviation of the mirror to mirror interval from the exact $T/2$ (if in a flat potential) for the wave-packets to overlap at the end, versus $T/4$. 
The experimental results shown in the graph are for static guide interferometers, for $T$ = 20, 120, 160, 200, and 220\,ms.
We fit the data to the function $2t(\omega t)^2$, where only $\omega$ is the free parameter. 
The fitted trap frequency is 0.51\,Hz. 

\end{document}